\documentclass[10pt,conference]{IEEEtran}
\usepackage{epsfig,setspace,amsmath,epsf,amssymb,bm,theorem,cite,graphicx, epstopdf,algorithm,algpseudocode,float,color,mathtools,authblk,tikz,physics}

\usetikzlibrary{positioning}
\usetikzlibrary{decorations.text}
\usetikzlibrary{decorations.pathmorphing}
\usepackage[short,c2]{optidef}

\usepackage{empheq}
\usepackage{cases}

\usepackage[table,xcdraw]{xcolor}
\usepackage{booktabs}
\usepackage{subfigure}
\usepackage{bbm}

\newtheorem{theorem}{Theorem}

\newtheorem{definition}{Definition}

\newtheorem{lemma}{Lemma}
\newtheorem{example}{Example}
\newenvironment{Proof}[1]{\medskip\par\noindent{\bf Proof:\,}\,#1}{{\mbox{\,$\blacksquare$}\par}}

\IEEEoverridecommandlockouts
\allowdisplaybreaks
\begin{document}
\title{Dual-Regime Absorbing Markov Chain Theory in Remote Estimation: Age-Minimizing Push Policies \thanks{I. Cosandal and S. Ulukus are with the University of Maryland, College Park, MD, USA. N. Akar is with Bilkent University, Ankara, T\"{u}rkiye. Corresponding author: S. Ulukus (email: ulukus@umd.edu).

This work is done when N.~Akar was on sabbatical leave as a visiting professor at the University of Maryland, MD, USA.}}

\author{Ismail Cosandal \qquad Sennur Ulukus \qquad Nail Akar}

\maketitle

\begin{abstract}
    For a remote estimation system, we study the optimization of age of incorrect information (AoII), which is a recently proposed semantic-aware information freshness metric. In particular, we assume an information source that observes a discrete-time finite-state Markov chain (DTMC), and occasionally transmits status update packets to a remote monitor which is tasked with remote estimation of the source. For the forward channel from the source to the monitor, we assume the channel delay to be modeled by a general discrete-time phase-type (DPH) distribution, whereas the reverse channel from the monitor to the source is assumed to be perfect, ensuring that the source has perfect information on the AoII and the remote estimate at the monitor, at all times. Push-based transmissions are initiated when AoII exceeds a threshold depending on the current estimation value, i.e., multi-threshold policy. In this very general setting, our goal is to minimize a weighted sum of the time average of a polynomial function of AoII, depending on the remote estimate, and energy consumption from transmissions. We formulate the problem as a semi-Markov decision process (SMDP) {\em with the same state-space of the original DTMC} to obtain the optimal multi-threshold policy, whereas the parameters of the SMDP are obtained by using a novel stochastic tool called \emph{dual-regime absorbing Markov chain} (DR-AMC), and its corresponding absorption time distribution named as \emph{dual-regime DPH} (DR-DPH). The proposed method is validated with numerical examples using comparisons against other policies obtained by exhaustive search, and also various benchmark policies. 
\end{abstract}

\section{Introduction}
Remote sensing systems have been receiving considerable attention due to technological advances making sensors more affordable and applicable \cite{bharathidasan2002sensor}. One of the main objectives in remote sensing systems is to keep the information fresh at the remote monitors. Recently, several freshness metrics have been proposed to quantify information freshness. The first of these metrics is the age of information (AoI) metric \cite{Yates__HowOftenShouldone, yates-survey} that quantifies information freshness by keeping track of how long ago the latest received information packet was generated. However, AoI may fall short of capturing freshness in certain estimation problems since it does not consider the dynamics of the sampled process \cite{maatouk2020}. Particularly, even when the latest received packet may have been generated a long time ago, it is possible that the source may not have changed since then, and therefore, the packet can still be fresh. Similarly, a recently received packet may contain stale information if the source has already changed its state after the packet was generated. Stemming from these drawbacks inherent to AoI, \cite{maatouk2020} proposes an alternative freshness metric, namely, age of incorrect information (AoII) that penalizes the mismatch between the source and its estimation over time, and regardless of when it is sampled, it defines the estimation as \emph{fresh} if it is the same as the source. Another prominent feature of AoII in contrast to AoI is that the monitor is not required to get a new sample to bring the age down to zero since the mismatch condition between the source and the monitor may as well be brought to an end upon a transition of the source to the estimated value at the monitor.

\begin{figure}[tb]
	\centering
    \includegraphics[width=0.9\columnwidth]{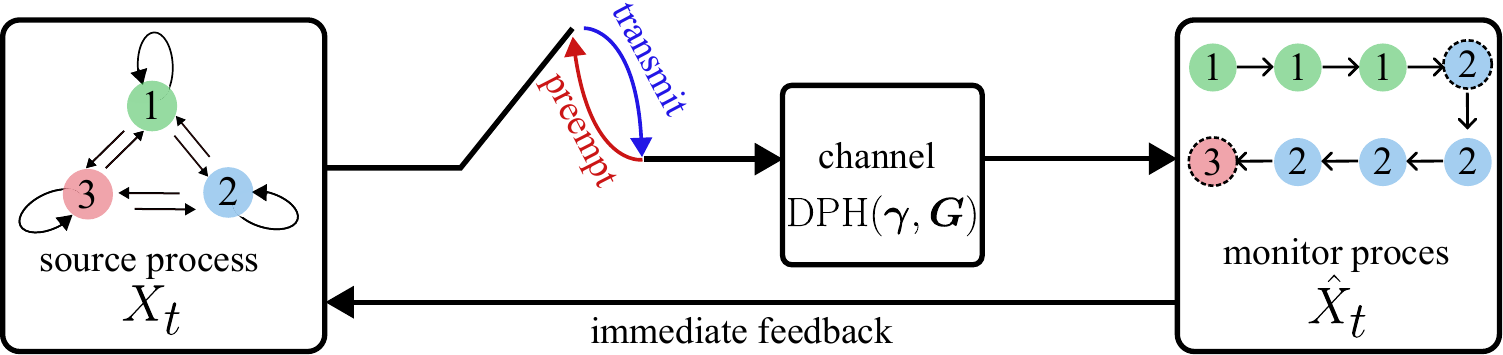}
	\caption{The remote estimation system involving the source process $X_t$, the monitor process $\hat{X}_t$, and the forward channel modeled by DPH($\bm{\gamma},\bm{G}$). The monitor updates its estimation with the received updates (marked with dashed circles).}
	\label{fig:SystemModel}
\end{figure}

Let us consider the remote estimation system in Fig.~\ref{fig:SystemModel}. For the source process $X_t$ and its remote estimation $\hat{X}_t$ at time $t$, the AoII process, denoted by
$\text{AoII}_t$, is given by,
\begin{align}
    \text{AoII}_t & = t-\sup \{t^\prime: t' \leq t, X_{t'}=\hat{X}_{t'} \} , \label{eq:AoII}    
\end{align}
which is in line with the original formulation in \cite{maatouk2020} for AoII, which considers a linear time penalty function with unit proportionality constant. In this paper, we focus on minimizing arbitrary functions of $\text{AoII}_t$ and estimation value $j$ which is denoted by $f_j(\text{AoII}_t)$, named as the AoII penalty functions. This dependence of the AoII function on the estimation value is motivated by applications that follow the classical \emph{missile detection example} in detection and estimation textbooks such as in \cite{poor2013introduction}, where incorrect estimation of the presence of a missile results in a higher cost than the incorrect estimation of its absence. A similar approach is used in \cite{xu2025timely}, where different age functions are used for nodes in a gossip network.
 
In the literature, it is proven that threshold-based policies, also known as switching-type policies, are optimal for a wide range of AoII problems. The works in \cite{maatouk2020, chen2022preempting, maatouk2022age, bountrogiannis2024age} consider symmetric discrete-time Markov chain (DTMC) sources, which are the most frequently studied type in the early AoII works, for which the optimum transmission policies are obtained in order to minimize the average AoII with and without energy constraints. These works propose the optimum policy in the form of a single threshold, and the optimization problem is cast as a Markov decision process (MDP). On the other hand, in \cite{cosandal2024modelingC}, a general continuous-time Markov chain (CTMC) source is studied for the first time, to the best of our knowledge. In \cite{cosandal_etal_TRIT24}, it is shown that using a single threshold push-based AoII policy is suboptimal for asymmetric sources, and it is proven that the optimal policy is a multi-threshold policy for which the threshold values depend on the states of both the source and estimation processes. In addition, the optimality of a similar threshold structure is derived for energy harvesting problems in \cite{zakeri2024semantic, arafa2019age, arafa2021timely}, where the thresholds also depend on the battery level. Threshold-based policies are also proposed for the AoI metric \cite{kahraman2023age} in different settings, and the metrics derived from AoII \cite{luo2025cost, joshi2021minimization, schiavo2025bidirectional}.

An absorbing Markov chain (AMC) is a Markov chain that has a number of transient and absorbing states, and the process starts operation in a transient state and evolves until an absorbing state is reached, upon which the process is said to be absorbed (or gets stuck) \cite{kemeny1960finite}. The distribution of the time until absorption for an absorbing Markov chain is known as a phase-type (PH) distribution, or more specifically \emph{discrete-time PH} (DPH) for discrete-time AMCs \cite{telek_book}. In this paper, we use the DPH distribution to model the general forward channel in Fig.~\ref{fig:SystemModel}. DPH distributions have recently been used for information freshness and networked control problems in several existing works. In \cite{akar2023distribution, akar2024age, gursoyenergy},  distributions of AoI and peak AoI processes are derived by making use of AMC and PH distributions. These works can be considered as an alternative to the stochastic hybrid systems (SHS) approach that is widely used to find the distribution of AoI \cite{yates2020age, moltafet_etal_tcom22}. In another work \cite{scheuvens2021state}, PH distributions are used to find the expected time before a certain number of consecutive packet failures occur in a wireless closed-loop control system. In this paper, we extend the well-established concepts of AMC and DPH, to the case where transient states and transition probabilities from the transient states are different depending on whether the elapsed time of the AMC is below or above a threshold, namely dual-regime AMC (DR-AMC) and dual-regime DPH (DR-DPH), respectively. A similar approach is used for threshold-based server selection problem involving the AoI metric in \cite{akar_Asilomar}, which involves more than two regimes. DR-AMC and DR-DPH have the potential to be used for analysis and optimization of threshold policies involving information freshness. 

In \cite{cosandal_etal_TRIT24}, it was shown that finding the optimum state-and-estimation based multi-threshold policy that minimizes the average AoII under an energy consumption constraint requires computation with complexity $\mathcal{O}(N^6)$, for a CTMC process with $N$ states. Thus, such a multi-threshold policy may not be suitable for large $N$. On the other hand, a relaxed policy in which the threshold values only depend on the estimation process results in similar performance in comparison to the optimal policy in most cases with relatively lower complexity, i.e., $\mathcal{O}(N^3)$. Therefore, in this paper, we focus our attention to the estimation-based multi-threshold policy in which the source initiates the transmission only if the duration of the mismatch between the source and the estimation processes exceeds a threshold $\tau_j$ when the estimation is $\hat{X}_t=j$, and we seek the optimum thresholds which minimize a weighted sum of the AoII cost and the consumed energy for transmissions, referred to as the transmission cost. We first formulate the problem as an SMDP \cite{white1993mdp} where the states are the embedded values at the embedded synchronization points, and the duration between two successive embedded points is random. Subsequently, we employ the DR-AMC theory to obtain the parameters of the SMDP. Finally, we obtain the optimum policy by the policy iteration algorithm. 

The contribution of our paper can be summarized as follows:
\begin{itemize}
    \item We study the (unconstrained) AoII and transmission cost minimization problem under an estimation-based AoII penalty function, using a general DPH-distributed forward channel model.
    \item We reduce the four-dimensional joint process composed of the state, estimation, channel phase, and age processes, to a one-dimensional embedded DTMC, which enables us to formulate the problem as an SMDP with reasonable complexity, and to obtain optimal push-based sampling policies using multiple thresholds, in an efficient manner.
    \item We propose the novel DR-AMC and DR-DPH analytical frameworks to obtain the distribution of AoII throughout the duration between two successive embedded points, along with the transition probabilities at two successive embedded time points. Hence, we can calculate the average of an arbitrary function of AoII numerically, which is key to the SMDP formulation. Furthermore, we obtain closed-form expressions for the same quantity for the special case of polynomial functions of AoII.
    \item We additionally propose a mixture policy to adapt the unconstrained optimization problem to the energy-constrained one.
\end{itemize}

The organization of the paper is as follows. In Section~\ref{sec:prel}, we present preliminaries on notation, and the AMC, DPH and SMDP frameworks, which are needed for the development of the paper. In Section~\ref{sec:dr-amc}, we present the mathematical frameworks of DR-AMC and DR-DPH that we introduce in this paper. The system model is described in Section~\ref{sec:systemmodel}. The optimization problem is formulated as an SMDP in Section~\ref{sec:embedded} through an embedded DTMC whose parameters are derived in Section~\ref{sec:calculation}. We dedicate Section~\ref{sec:numerical} to the numerical results, and finally, we conclude in Section~\ref{sec:conclusions}.

\section{Preliminaries} \label{sec:prel}

\subsection{Notation}
Throughout the paper, we use lowercase and uppercase bold characters for a vector, and a matrix, respectively. Specifically, ${a}_{m}$ denotes the $m$th entry of the vector $\bm{a}$, and $a_{mn}$ denotes the $(m,n)$th element of the matrix $\bm{A}$. The $N\times N$ identity matrix is denoted by $\bm{I}_N$, but the subscript can be omitted for convenience when the size of the matrix can be inferred.
A $\bm{1}$ denotes a column vector of ones, $\bm{e}_k$ denotes a column vector of zeros except for the $k$th entry, which is one. Finally, the operation $\otimes$ corresponds to the Kronecker product \cite{linearAlgebra}.

\subsection{Absorbing Markov Chains and Phase-Type Distribution}
The AMC we study in this paper refers to a DTMC that has $K$ transient and $L$ absorbing states \cite{latouche1999introduction}. The process starts from a transient state (or phase) and evolves until reaching one of the absorbing states. Consider the AMC $Y_t\in\{1,2,\dots K, K+1,\ldots, K+L\}$, $t=0,1,\dots$, where the first $K$ states are the transient states, and the last $L$ states are the absorbing states. The transition matrix of this process can be written as 
\begin{align}
    \bm{Q}=
    \begin{bmatrix}
            \bm{A} & \bm{B} \\ \bm{0} & \bm{0}
    \end{bmatrix},
\end{align}
where $\bm A_{K \times K}$ and $\bm B_{K \times L}$ are the transient probability transition sub-matrix (TPTS) and absorption probability transition sub-matrix (APTS) corresponding to the transition probabilities among the transient states, and from the transient states to the absorbing states, respectively. In this case, we say that $Y_t$ is an AMC characterized with the triple $(\bm{\beta},\bm A,\bm B)$, i.e., $Y_t \sim \text{AMC}(\bm{\beta},\bm A,\bm B)$, where $\bm{\beta} = \{ \beta_i \}$ is the ${1 \times K}$ initial probability vector (IPV), and
\begin{align}
    \beta_i=\mathbb{P}(Y_0=i), \quad i= 1,\dots,K.
\end{align}
We write the transient probability vector of the AMC $Y_t$ of size $1 \times K$ at time $k$ as follows,
\begin{align}
   \bm{y}_t & = 
   \begin{pmatrix}
      y_{t,1} & y_{t,2} & \cdots & y_{t,K}  
   \end{pmatrix}, \quad
   y_{t,k}   = \mathbb{P} (Y_t=k), \label{eq:yki}
\end{align} 
which leads to the following closed-form expression for $\bm{y}_t$,
\begin{align} 
    \bm{y}_t & = \bm{\beta} \bm{A}^{t}, \quad t \geq 1. 
\end{align} 
A basic property about an AMC is that the expected number of visits to a transient state $j$ starting from a transient state $i$ is given by the $(i,j)$th entry of the fundamental matrix \cite{kemeny1960finite}
\begin{align}
    \bm F=(\bm I-\bm A)^{-1}. \label{eq:F}
\end{align}
We denote the time until absorption by $T$, which corresponds to the first time slot that the process is in any absorbing state, and mathematically it can be expressed as
\begin{align}
    T=\max(t|X_{t-1}\in\{1,\dots,K\}, X_{t}\in\{K+1,\dots,K+L\}).
\end{align}
Upon merging all the absorbing states into one, the distribution of $T$ is known as the DPH distribution \cite{latouche1999introduction}, i.e., $T\sim\text{DPH}(\bm{\beta},\bm{A})$. The absorption time $T$ has  cumulative distribution function (cdf) $F_T(t)= \mathbb{P} (T \leq t)$, and probability mass function (pmf) $p_T(t)=\mathbb{P} (T =t)$, which can respectively be written as follows for any general absorption time $t$,  
\begin{align}
    F_T(t) & = 1 - \bm{y}_t \bm{1} = 1 - \bm{\beta} \bm{A}^t \bm{1}, \label{eq:cdf}\\
    p_T(t) & = F_T(t) - F_T(t-1) = \bm{\beta} \bm{A}^{t-1} (\bm{1}-\bm{A}\bm{1}). \label{eq:pmf}
\end{align}
Several well-known distributions with finite/infinite support can be represented by DPH distributions; see Appendix~\ref{app:channel} for various examples. The following definition is needed for the factorial moments.

\begin{definition}[Falling factorial power \cite{graham1994concrete}]
    The falling factorial power of $x$ of order $m$, also called \emph{$x$ to the $m$ falling}, denoted by $x^{\underline{m}}$, is given by
    \begin{align}
        x^{\underline{m}}=x(x-1)(x-2) \cdots (x-m+1).
    \end{align}
\end{definition}

The $i{\text{th}}$ factorial moment of $T$, denoted by $\nu_T(i)=\mathbb{E}[T^{\underline{i}}]$, can be written in closed form through the following expression \cite{telek_book},
\begin{align}
   \nu_T(i) & = \mathbb{E}[\, T  (T-1) \ \cdots \ (T-i+1) \,],  \\ 
   & = i! \, \bm{\beta} (\bm{I}-\bm{A})^{-i} \bm{A}^{i-1} \bm{1}.
   \label{factorialMoments}
\end{align}

\subsection{Semi-Markov Decision Process (SMDP)} \label{sec:smdp}
This section describes the average-cost SMDP framework, and presents the policy iteration method based on \cite{book_ross,ibe2013markov}. The discrete-time SMDP of interest to the current paper is the tuple $(\mathcal{S},\mathcal{A},\rho,r,d)$ described below.
\begin{itemize}
    \item The {\em state space} $\mathcal{S} = \{ 1,2,\ldots,S \}$ of the SMDP is the finite set of states that are visited by the SMDP process.
    \item The {\em action space} $\mathcal{A}$ is the set of actions that can be taken at a state. 
    The use of different action sets at different states is not of interest to this paper, and is therefore not discussed.
    \item $\rho: \mathcal{S} \times \mathcal{A} \times \mathcal{S} \rightarrow [0, 1]$  is called the {\em transition function} of the SMDP. Particularly, when action $a \in A$ is taken at current state $s$, a transition to state $s'$ occurs with probability $\rho(s,a,s')$.
    \item $r: \mathcal{S} \times \mathcal{A} \rightarrow [0, \infty)$ is the {\em cost function} which represents the average accumulated cost $r(s,a)$ when action $a$ is taken at state $s$, until the transition to the next state.
    \item $d: \mathcal{S} \times \mathcal{A} \rightarrow (0, \infty)$ is the {\em sojourn time function} 
    which is the average time spent at state $s$ when action $a$ is taken, which is denoted by $d(s,a)$.
\end{itemize}
A deterministic policy $\phi : S \rightarrow A$ is one that maps each state $s \in S$ to a single action $a \in A$, i.e., we take the action $\phi_s$ when we are at state $s$.
Let a deterministic policy $\phi$ be given. Let the embedded (at the decision epochs) Markov chain associated with the policy $\phi$  be denoted by $Z_t^\phi$, $t \geq 0$ where $Z_t^{\phi}$ is the state of the system at decision epoch $t$. Let $B_t^\phi$ denote the total cost accumulated up to time $t$, $t \geq 0$. If the embedded process $Z_t^\phi$ has no two disjoint communicating classes, then for each initial state $Z_0^\phi=s$, the limit 
\begin{align}
    \lim_{t \rightarrow \infty} \frac{B_t^{\phi}}{t} = r^{\phi}, \label{smdp}
\end{align}
exists, and is independent of the initial state $s$. The deterministic policy $\phi^*$ that minimizes the long-run average cost in \eqref{smdp} is the optimum solution for the SMDP problem. Algortithm~\ref{alg:SMDP} presents the pseudo-code for the policy iteration algorithm to obtain $\phi^*$ given the average-cost SMDP described by the tuple $(\mathcal{S},\mathcal{A},\rho,r,d)$ \cite{book_ross,book_tijms,ibe2013markov}. 

\begin{algorithm}[tbh]
    \caption{Policy Iteration Algorithm for an SMDP}\label{alg:SMDP}
    \begin{algorithmic}
    \State \textbf{Input:} Initiate all actions  $u_j$ $j \in \mathcal{N}$ with an arbitrary policy.
    \State \textbf{Output:} Policy $\phi^*$
    \State \textbf{Step 1: (Initialization)} Start with an initial arbitrary policy $\phi$
    \State \textbf{Step 2: (Relative Value Determination)} For the present policy $\phi$, solve the following linear system of equations for $s=1,2,\ldots,S-1$, for the relative values $V_s$ and the long-run average cost $r^{\phi}$ of the policy $\phi$,
    \begin{align}
                  r^{\phi} d(s,\phi_s) + V_s = r(s,\phi_s) + \sum_{s'=1}^S \rho(s,\phi_s,s') V_{s'},    
    \end{align}
    by setting $V_S=0$.
    \State{\bf Step 3: (Policy Improvement)} For each state $s=1,2,\ldots,S$, set the alternative action $a_s$ to,
         \begin{align}
                     \underset{a\in\mathcal{A}}{\text{arg min}} \ \ {\frac{1}{d(s,a)} \left( r(s,a) + \sum_{s'=1}^S \rho(s,a,s') V_{s'} - V_s \right)}  \label{eq:polimp}
     \end{align}
    \State {\bf Step 4:} Update the policy $\phi_s=a_s$ for $s=1,2,\ldots,S$
    \State {\bf Step 5:} Stop when the two successive policies are the same and set $\phi^*=\phi$. Otherwise, go to Step 2.   
\end{algorithmic}
\end{algorithm}

\section{Dual-Regime Absorbing Markov Chains and Phase-type Distributions} \label{sec:dr-amc}
In this section, we introduce the analytical frameworks of DR-AMC and DR-DPH, which are the extensions of AMC and DPH, respectively, for which the TPTS, APTS, and the number of transient states depend on the regime associated with the elapsed time since the AMC starts evolution. More specifically, we define a process $Y_t$ that, when the elapsed time is strictly below a threshold $\tau$, corresponds to the first $\tau-1$ time steps with time indices $t=(0,\dots,\tau-2)$, it is considered in \emph{regime 1}. During this regime, it has $K_1$ transient states and $L$ absorbing states. We denote the TPTS and APTS for this regime by $\bm{A}_1$ and $\bm{B}_1$, respectively, and the process starts from regime 1 with IPV $\bm{\beta}_1$. Therefore, the process evolves from time indices $t=(0,\dots,\tau-2)$ to $t'=(1,\dots,\tau-1)$ with $\bm{A}_1$ and $\bm{B}_1$, thus if it is absorbed in regime 1, the absorption time is $T<\tau$. On the other hand, if the process is not absorbed in regime 1, \emph{regime 2} starts from the $\tau$th time slot (at $t=\tau-1$). Regime $2$ has $K_2$ transient states in addition to the same $L$ absorbing states. Also, we define the \emph{intermediate transition matrix} (ITM) denoted by $\bm{\Theta}$, a $K_1 \times K_2$ matrix, corresponding to the transition probabilities among the transient states of regimes 1 and 2, at the regime cross-over instance. Mathematically, the IPV of regime 2 is obtained by
\begin{align}
    \bm{\beta}_2=\bm{\beta}_1\bm{A}^{\tau-1}\bm{\Theta}. \label{eq:ThetaEqn}
\end{align}
Similar to regime 1, we denote the TPTS and APTS for regime 2 by $\bm{A}_2$ and $\bm{B}_2$, respectively, and $T\geq\tau$ indicates it is absorbed in regime 2. Finally, we characterize the DR-AMC with the 7-tuple,
\begin{align}
   Y_t & \sim \text{DR-AMC} \left( \bm{\beta}_1,\tau,\bm{\Theta},\bm{A}_1,\bm{A}_1,\bm{B}_1,\bm{B}_2 \right). \label{eq:tuple_AMC}
\end{align}
Next, we define the DR-DPH distribution that corresponds to the distribution of the absorption time $T$ of the DR-AMC described by \eqref{eq:tuple_AMC}, which is obtained by means of merging the absorbing states into one. Hence, $T$ is
is characterized by the following 5-tuple,
\begin{align}
   T& \sim \text{DR-DPH} \left( \bm{\beta}_1,\tau,\bm{\Theta},\bm{A}_1,\bm{A}_2 \right). \label{eq:tuple_DPH}
\end{align}

\begin{theorem} \label{lem:drph}
    The absorption time $T$ of the DR-AMC $Y_t$ whose characterization is given in \eqref{eq:tuple_DPH} has the following pmf,
    \begin{align}
        p_T(t)& = \mathbb{P}(T=t) = \begin{cases}
            \bm{\beta}_{1}\bm{A}_{1}^{t-1}(\bm{1}-\bm{A}_{1}\bm{1}), &
         t<\tau, \\
         \bm{\beta}_{2}\bm{A}_{2}^{t-\tau}(\bm{1}-\bm{A}_{2}\bm{1}),
        & t\geq\tau, \label{eq:pt2}
        \end{cases}
    \end{align}
    where $\bm{\beta}_{2}$ is given in  \eqref{eq:ThetaEqn}.
\end{theorem}

\begin{Proof}
    Let us define the two random variables 
    \begin{align}
        T_1 & \sim \text{DPH}(\bm{\beta}_{1},\bm{A}_{1}), \quad T_2\sim \text{DPH}(\hat{\bm{\beta}}_{2},\bm{A}_{2}),
    \end{align}
    where $\hat{\bm{\beta}}_{2}$ is the IPV vector for the second regime conditioned on entrance to the second regime. Then, $\hat{\bm{\beta}}_{2}=\frac{1}{\mathbb{P}(T_{1}\geq\tau)}\bm{\beta}_{2}$. First, it is straightforward to see that the conditional distributions $\{T|t<\tau\}$ and $\{T_1|t<\tau\}$ are equivalent to each other, which proves the first case of \eqref{eq:pt2} using the expression for the pmf of the DPH distribution in \eqref{eq:pmf}. The second case of \eqref{eq:pt2} is equivalent to the probability that the AMC is not absorbed in the first regime, i.e., $T_1\geq\tau$, and it spends $t-\tau+1$ units of time without absorption in the second regime, which can be expressed as,
    \begin{align}
       \mathbb{P}(T_1\geq\tau,T_2=t-\tau+1)&=\mathbb{P}(T_1\geq\tau)\mathbb{P}(T_2=t-\tau+1) \nonumber \\
        &= \mathbb{P}(T_1>\tau)\hat{\bm{\beta}}_2\bm{A}_2^{t-\tau}(\bm{1}-\bm{A}_2\bm{1}) \nonumber\\
        &= \bm{\beta}_2\bm{A}_2^{t-\tau}(\bm{1}-\bm{A}_2\bm{1}),
    \end{align}
    which completes the proof.
\end{Proof}

We also define the absorption vector in regime $i$, for $i=1,2$, denoted by $\bm{\sigma}_i$, 
\begin{align}
   \bm{\sigma}_i & = \begin{pmatrix}
      \sigma_{i1} & \sigma_{i2} & \cdots & \sigma_{iL}  
   \end{pmatrix}, 
\end{align}
where $\sigma_{ij}$ is the probability of absorption into absorbing state-$j$ stemming from a transition taking place in regime $i$. Notice that  $\bm{A}_i$, $i=1,2$ are sub-stochastic matrices with all their eigenvalues being strictly inside the unit circle, thus the matrix $(\bm{I} - \bm{A}_i)$ is invertible for $i=1,2$ which ensures that the per-regime absorption vectors are non-zero. 

\begin{lemma}
    \label{lem:sigma}
    For $Y_t \sim \text{DR-AMC} \left( \bm{\beta}_1,\tau,\bm{\Theta},\bm{A}_1,\bm{A}_1,\bm{B}_1,\bm{B}_2 \right)$, the two per-regime absorption vectors $\bm{\sigma}_1$ and $\bm{\sigma}_2$  are written in closed form as,
    \begin{align}
        \bm{\sigma}_{1}&=\bm{\beta}_1 (\bm{I}-\bm{A}_1^{\tau-1})(\bm{I}-\bm{A}_1)^{-1}\bm{B}_1,  \label{eq:prob_dr1} \\
        \bm{\sigma}_{2}&=\bm{\beta}_2 (\bm{I}-\bm{A}_2)^{-1}\bm{B}_2.  \label{eq:prob_dr2}
    \end{align}
\end{lemma}

\begin{Proof}
    We first define the absorption probability vector $\bm{\tilde{y}}_t$, similar to the definition of $\bm{y}_t$ in \eqref{eq:yki}, at time $t$ as,
    \begin{align}
        \bm{\tilde{y}}_t & = \begin{pmatrix}
          \tilde{y}_{t,l} & \cdots & \tilde{y}_{t,L}  
       \end{pmatrix}, \\
       \tilde{y}_{t,l}   &= \mathbb{P} (Y_t=K_i+l), \quad i=1,2. \label{yk}
    \end{align} 
    Let us first focus on the transitions in regime $1$, we have,
    \begin{align}
        \begin{pmatrix}
          \bm{y}_t & \bm{\tilde{y}}_t 
        \end{pmatrix}  & = \begin{pmatrix}
          \bm{\beta}_1 & \bm{0}
        \end{pmatrix} \left( \begin{array}{c|c} \bm{A}_1 & \bm{B}_1 \\ \hline \bm{0} & \bm{I} \end{array} \right)^k, \  0 \leq k \leq \tau_1.
    \end{align}
    Therefore, for $0 \leq t < \tau$,
    \begin{align}
        \bm{y}_t & = \bm{\beta}_1 \bm{A}_1^{t}, \label{xkR1} \\  
        \bm{\tilde{y}}_t & = \bm{\beta}_1 \sum_{l=0}^{t-1}  \bm{A}_1^{l}\bm{B}_1,  \\ 
        & = \bm{\beta}_1   (\bm{I} - \bm{A}_1^t) (\bm{I} - \bm{A}_1)^{-1} \bm{B}_1.
        \label{ykR1} 
    \end{align}
    Since $\tilde{y}_{\tau-1,j}$ is the probability that absorption occurs into absorbing state-$j$ from a transition in regime $1$, we have
    \begin{align}
        \bm{\sigma}_{1} & = \bm{\tilde{y}}_{\tau-1}, \label{gammaR1ilk}\\
        & = \bm{\beta}_1  (\bm{I} - \bm{A}_1^{\tau-1})  (\bm{I} - \bm{A}_1)^{-1} \bm{B}_1  . \label{gammaR1}
    \end{align}
    Similarly, for regime $2$, we can express $\bm{y}_t$, $t\geq\tau$ as
    \begin{align}
        \bm{y}_t & = \bm{\beta}_2 \bm{A}_2^{t-\tau}, \label{xkR2} \\  
        \bm{\tilde{y}}_{t} & = \bm{\sigma}_1 + \bm{\beta}_2 \sum_{l=0}^{t-1}  \bm{A}_2^{l}\bm{B}_2,  \\ 
        & = \bm{\sigma}_1 + \bm{\beta}_2 (\bm{I} - \bm{A}_2^t)  (\bm{I} - \bm{A}_2)^{-1} \bm{B}_2.
        \label{ykR2} 
    \end{align}
    Notice that $\lim_{l \rightarrow \infty}\bm{A}_2^l = \bm{0}$ since $\bm{A}_2$ is a sub-stochastic matrix. Subsequently, we can write 
    \begin{align}
        \lim_{t \rightarrow \infty} \bm{y}_t = \bm{\sigma}_1+ \bm{\beta}_2 (\bm{I} - \bm{A}_2)^{-1} \bm{B}_2, \label{temp347}
    \end{align}
    which gives the absorption probabilities in infinite horizon, i.e., from regime $1$ or $2$. Finally, we write,
    \begin{align}
        \sigma_{2} & = \bm{\beta}_2 (\bm{I} - \bm{A}_2)^{-1} \bm{B}_2,
    \end{align}
    by extracting from \eqref{temp347} the absorption probabilities from regime $1$.
\end{Proof} 

\begin{lemma} \label{lem:pow}
    The $m{\text{th}}$ factorial moment of $T$, denoted by $\nu_T(m)=\mathbb{E}[T^{\underline{m}}]$, is given in \eqref{eq:pow} located at the top of the next page. Additionally, the ordinary moments, denoted by $\mu_T(m)=\mathbb{E}[T^{{m}}]$, can be obtained from the corresponding factorial moments as follows,
    \begin{align}
        \mu_T(m)=\sum_{r=0}^m S(m,r) \nu_T(r), \label{eq:mom}
    \end{align}
    where $S(m,k)$ is the Stirling number of the second kind.
\end{lemma}

The proof of Lemma~\ref{lem:pow} is given in Appendix~\ref{app:fact}.

\begin{figure*}
        \begin{align}
        \nu_T(m)=\begin{cases}
            \bm{\beta}_1 \sum_{t=1}^{\tau-1} t^{\underline{m}}\bm{A}_1^{t-1}(\bm{1}-\bm{A}_1\bm{1})+\bm{\beta}_2 \sum_{r=0}^m \binom{m}{r} \tau^{\underline{m-r}}r!\bm{A}_2^r (\bm{I}-\bm{A}_2)^{-r-1}(\bm{1}-\bm{A}_2\bm{1}),
         & m\leq \tau, \\
            \bm{\beta}_1 \sum_{t=1}^{\tau-1} t^{\underline{m}}\bm{A}_1^{t-1}(\bm{1}-\bm{A}_1\bm{1})+ \bm{\beta}_2 \sum_{r=0}^m \binom{m}{r} m^{\underline{m-r}}r!\bm{A}_2^r (\bm{I}-\bm{A}_2)^{-r-1}   \bm{A}_2^{m-\tau}(\bm{1}-\bm{A}_2\bm{1}),
         & m> \tau,             \end{cases} \label{eq:pow}
    \end{align} 
\end{figure*}

\begin{example}
In this illustrative example, we formulate the distribution of a generic age process $\Delta_t$ with a control variable $U_t$. We consider that the age process evolves based on the control variable $U_t$ as
\begin{align}
    \Delta_{t}=\begin{cases}
        \Delta_{t-1}+1, & U_t=0, \\
        0, & U_t=1.
    \end{cases}
\end{align}
Furthermore, we assume that transition probabilities of $U_t$ change based on the age process $\Delta_t$, which is strictly less than a threshold $\tau$ or not. For this example, we consider when $\Delta_t<\tau$, the value of $U_t$ changes from $0$ to $1$ with probability $q$; mathematically its transition probability is expressed as
\begin{align}
    \mathbb{P}(U_{t+1}=1|U_t=0,\Delta_t<\tau)=q.
\end{align}
In addition, when $\Delta_t\geq\tau$, the control variable reaches state $1$ in $D$ steps, where $D$ is modeled with a mixture of geometric distributions, and it is denoted as MGD$(p_1,p_2,w_1,w_2)$. It can be considered as a communications problem under random channel conditions, where the channel is either in a good condition with probability $w_1$, and the transmission duration is distributed with Geo$(p_1)$, or in a bad condition with probability $w_2$, and the transmission duration is distributed with Geo$(p_2)$. For this example, we define the event $U_t=1$ as the absorbing state, and duration until the absorption starting from $(U_0=0, \Delta_0=1)$ can be modeled as $\text{DR-DPH} \left( \bm{\beta}_1,\tau,\bm{\Theta},\bm{A}_1,\bm{A}_2 \right)$ with the following formulation:

The first regime includes a single transient state that is responsible for continuing $U_t=0$, and as a natural result $\bm{\beta}_1=1$. During the first regime, the absorption occurs with probability $q$, or it stays in the transient state with probability $1-q$. Therefore, TPTS and APTS of this regime are 
\begin{align}
    \bm{A}_1=\begin{bmatrix}
        1-q
    \end{bmatrix}, \quad 
    \bm{B}_1=\begin{bmatrix}
        q 
    \end{bmatrix}.
\end{align}
At the ($\tau-1$)th slot, the age value becomes $\tau$, and the regime $2$ starts. Transient states of regime $2$ correspond to the channel condition. The probability of reaching this state is $\bm{\beta}_1\bm{A}_1^{\tau-1}=(1-q)^{\tau-1}$, and by multiplying the ITM, corresponding to the probability of the channel conditions,
\begin{align}
  \bm{\Theta}=\begin{bmatrix}
    w_1 & w_2
\end{bmatrix},  
\end{align}
we get the IPV of the second regime as
\begin{align}
    \bm{\beta}_2=\bm{\beta}_1\bm{A}_1^{\tau-1}\bm{\Theta}=\begin{bmatrix}
    w_1(1-q)^{\tau-1} & w_2(1-q)^{\tau-1}
\end{bmatrix}
\end{align}

At any time slot of the second regime, the process continues until the absorption with probability $p_i$ when the channel condition is $i=1,2$. Therefore, we can express TPTS and APTS of the second regime as
\begin{align}
    \bm{A}_1=\begin{bmatrix}
        1-p_1 & 0 \\ 0 & 1-p_2
    \end{bmatrix}, \quad 
    \bm{B}_1=\begin{bmatrix}
        p_1 \\ p_2
    \end{bmatrix},
\end{align}
which finalizes the formulation.
\end{example}

We refer the reader to \cite{cosandal_etal_TRIT24} and \cite{akar_Asilomar} for further extensions of DR-AMC and DR-DPH to the case of more than two regimes, namely multi-regime extensions, in continuous-time and discrete-time, respectively.

\section{System Model} \label{sec:systemmodel}
We consider the time-slotted remote estimation system in Fig.~\ref{fig:SystemModel} with an $N$-state DTMC information source process $X_t\in \mathcal{N}=\{1,2,\dots,N\}$ which has an irreducible transition probability matrix $\bm{Q}$, and the one-step transition probability from state $i$ to state $j$ is denoted by $q_{ij}$. With the generate-at-will (GAW) principle, the source can initiate a transmission of a status update packet carrying the observation value to the remote monitor at the beginning of a time slot. We assume the events happen in the following order: i) the state of $X_t$ changes just before the end of the time slot, and ii) an ongoing transmission is only completed at the end of the time slot. Therefore, if the source process changes when there is an ongoing transmission, the source preempts the ongoing transmission with a fresh packet, to avoid sending \emph{incorrect information}. The ongoing transmission continues until the source state changes, or the transmission is successful. We assume instantaneous feedback from the monitor to the source. Therefore, the source is always aware of the estimation and AoII processes. Consequently, we propose an estimation-based multi-threshold transmission policy for which the source always initiates a new transmission, or continues thr ongoing transmission when AoII value exceeds the threshold $\tau_j$ when the estimation $\hat{X}_t=j$. 

We model the channel delay $D$ to be distributed according to a DPH distribution, $D \sim (\bm{\gamma},\bm{G})$, with $M$ transient states, called \emph{channel phases}, and a single absorption state corresponding to successful transmission. When the transmission of a packet is initiated, the channel is in phase $m$ with probability $\gamma_m$. If transmission is not preempted, or equivalently if the source stays the same, until the next time slot, the channel evolves to phase $m'$ which occurs with probability $g_{mm'}$, or the transmission is complete with probability $h_m$. Otherwise, transmission of a new packet is initiated when the channel is in phase $m$ with probability $\gamma_m$.

As the estimator at the monitor, we employ the martingale estimator which uses the latest received information as its estimate, i.e., $\hat{X}_t=X_{t'}$, where $t'$ is the generation time of the latest successful transmission \cite{akar_ulukus_tcom24}. Hence, in the martingale estimator, the estimate can only be updated at packet reception instances. This estimation rule is widely used in the literature due to its amenability to analysis and optimization. However, we refer to \cite{cosandal2024joint, cosandal2025sensor} which additionally considers the maximum a posteriori (MAP) estimator in pull-based remote estimation systems for which the monitor can update its estimation as the maximum likely state, without having to receive a status update.

\begin{figure}[t]
    \centering
    \vspace{0.05in}
    \includegraphics[width=0.95\linewidth]{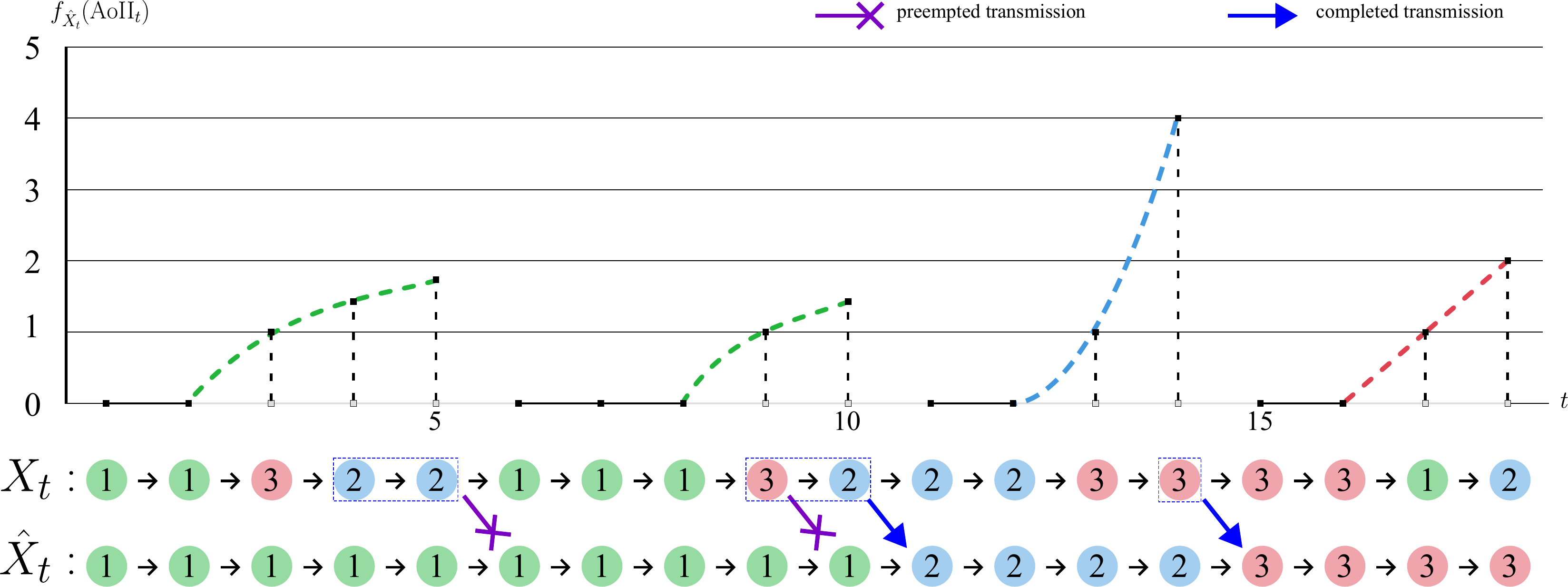}
    \caption{Sample path of the processes $X_t$, $\hat{X}_t$, and AoII$_t$ for a general transmission policy, and AoII penalty functions $f_1(x)=\sqrt{x}$, $f_2(x)=x^2$, and $f_3(x)=x$, where $x = \text{AoII}_t$. Successful transmissions are shown with blue arrows, and purple arrows indicate preempted transmissions. Time slots when the source is transmitting are illustrated with dashed boxes. AoII is reset upon synchronization, otherwise, it increases by one at every time slot as long as the mismatch condition between $X_t$ and $\hat{X}_t$ remains.}
    \label{fig:main}
\end{figure}

The mismatch between $X_t$ and $\hat{X}_t$ is measured with the AoII process defined in \eqref{eq:AoII}, where $f_j(x)$ corresponds to the AoII penalty function for estimation $j$. In Fig.~\ref{fig:main}, an example scenario is given for $N=3$, and penalty functions $f_1(x)=\sqrt{x}$, $f_2(x)=x^2$, and $f_3(x)=x$, where $x=\text{AoII}_t$. We highlighted the time slots when there is an ongoing transmission with dashed boxes. Note that, transmissions at time slots $5$ and $9$ are preempted because of a state change in the next time slot. On the other hand, transmissions are successful at time slots $11$ and $15$, and they result in a change of the estimation process $\hat{X}_t$. Notice that, at time slot $6$, AoII is reset without a successful transmission since the original process $X_t$ transitions to the estimated value.
 
The first objective of this work is to find a transmission policy for the source that minimizes the average cost,  
\begin{mini}
	{\substack{\bm{\tau}\in \mathbb{Z}_{+}^{N}}}
	  {\lim_{L\to \infty}\frac{1}{L}\sum_{t=1}^L\text{cost}_t} 
	{\label{Opt1}}
    {}
\end{mini}
where $\bm{\tau}$ is the vector of threshold values, and
\begin{align}
    \text{cost}_t =f_{\hat{X}_{t}}(\text{AoII}_t)+\lambda\delta_t, \label{eq:cost}
\end{align}
where $\delta_t$ is one if there is a transmission at time $t$, and zero otherwise, and $\lambda$ denotes a relative weight assigned for the transmission cost. We also consider the following constrained optimization problem,
\begin{mini}
	{\substack{\bm{\tau}\in \mathbb{Z}_{+}^{N}}}
	 {\lim_{L\to \infty}\frac{1}{L}\sum_{t=1}^L f_{\hat{X}_{t}}(\text{AoII}_t)} 
	{\label{Opt2}}
    {}
    \addConstraint{ \lim_{L\to \infty}\frac{1}{L} \delta_t }{\leq \alpha,}
\end{mini}
where $\alpha$ denotes the sampling budget.

\section{Embedded DTMC Representation} \label{sec:embedded}
The conventional way to obtain the optimum policy is to define an MDP with a four-dimensional state-space, namely $(X_t, \ P_t,\ \hat{X}_t, \ \text{AoII}_t)$, which is detailed in Appendix~\ref{app:stand}. However, we take a different approach in this paper by obtaining an embedded DTMC whose state-space is the same as the original process $X_t$, as opposed to the four-dimensional state-space of Appendix~\ref{app:stand}. Subsequently, we formulate the minimization problem in \eqref{Opt1} as an SMDP. Additionally, with this approach, we avoid truncation of the state-space which would be required since $\text{AoII}_t$ can take arbitrarily large values. In order to obtain the embedded DTMC, we first need to define \emph{embedded points (EP).}

\begin{definition}[Embedded Point]
    A time point $t_0$ is called an embedded point (EP) with embedded value (EV) $E_j$ satisfying $\hat{X}_{t_0-1} \neq  {X}_{t_0-1}$, ${X}_{t_0} = \hat{X}_{t_0+1}$. Thus, embedded time points correspond to the time index when the source and monitor processes just get to synchronize at value $j$. 
\end{definition}

The interval between the EP with EV $E_j$ and the next EP is called a cycle of type $j$, or cycle-$j$ in short. Cycle-$j$ is divided into two separate intervals with the first one called the \emph{in-sync interval} with duration $H_j$ which starts from the value $E_j$, and lasts until the first time slot at which synchronization between the source and the monitor is broken. At this time instant, the second interval gets to start and lasts until the beginning of the next cycle, called the \emph{out-of-sync interval}, whose duration is denoted by $T_j$. During the out-of-sync interval, there is a transmission whenever AoII$_t$ exceeds the threshold $\tau_j$, which needs to be obtained for each $j$ to minimize the objective function in \eqref{Opt1}. Additionally, we denote the duration of cycle-$j$ by $D_j=H_j+T_j$, the total AoII cost by $A_j=\sum_{t=1}^{T_j} f_j(t)$, and the total number of slots used for transmissions in cycle-$j$ by $C_j$.

\begin{figure}[t]
    \centering
        \vspace{0.05in}
    \includegraphics[width=0.85\linewidth]{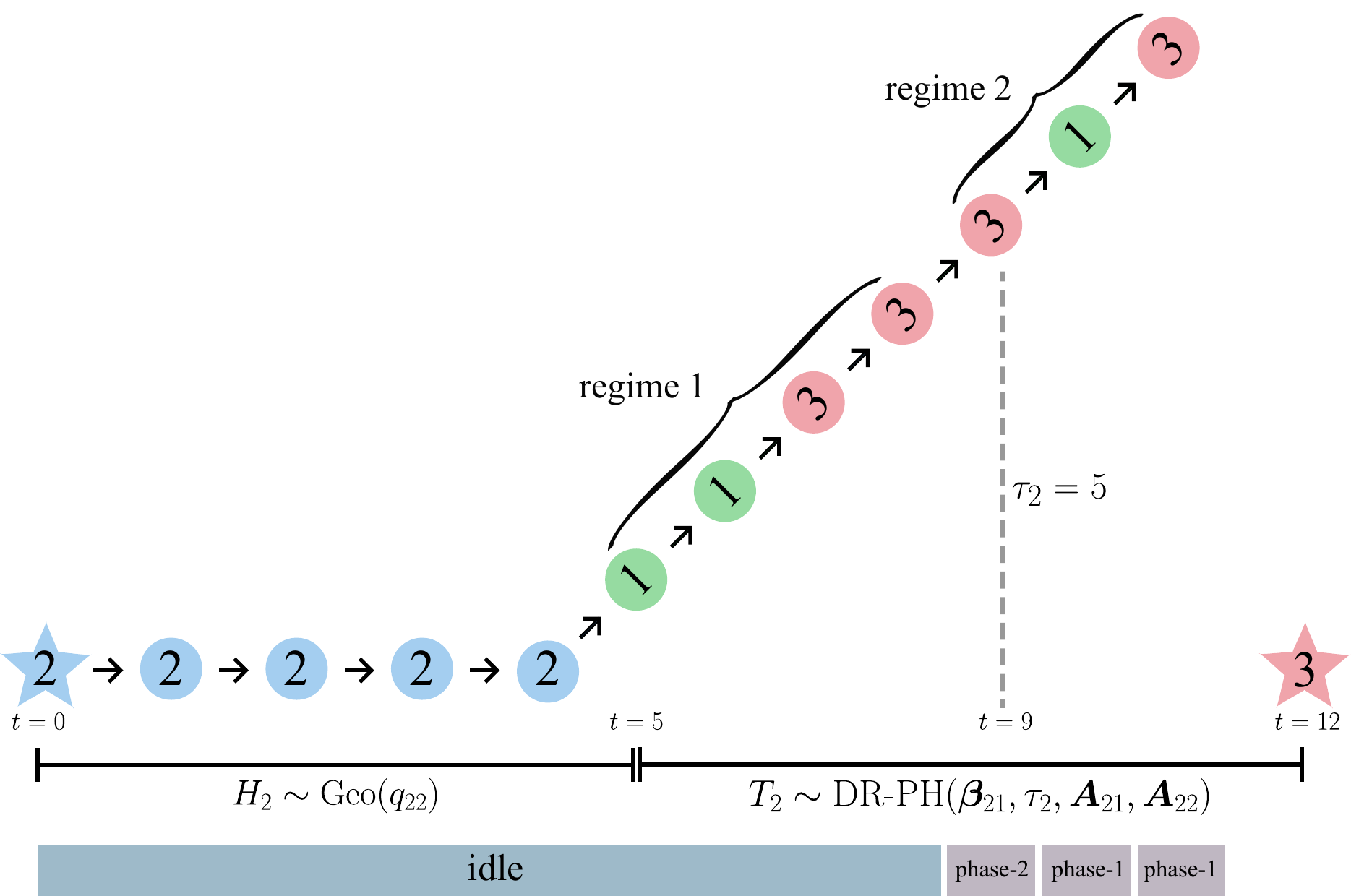}
    \caption{A sample path of the cycle-$2$ with $\tau_2=5$. Distributions of $H_2$ and $T_2$ are detailed in Example~\ref{ex:cycle2}.}
    \label{fig:SP}
\end{figure}

A sample path of cycle-$2$ is illustrated in Fig.~\ref{fig:SP}. At time $t=0$ cycle-$2$ starts from $E_2$, and the in-sync interval lasts for $H_2=5$ time slots. At time $t=5$, the synchronization is broken with a state change of the source, and the out-of-sync interval starts. The first $4$ time slots of this interval are the regime-$1$ when the source refrains from transmission. At time $t=9$, AoII$_t$ value reaches the threshold $\tau_2=5$ which initiates the regime $2$. There is an ongoing transmission from $t=9$ to $t=11$, and the corresponding cycle and the out-of-sync interval are completed with the completion of the transmission at the end of $t=11$. In this example, the total duration of the cycle is $D_2=H_2+T_2=12$, the total transmission cost is $C_2=3$, and the total AoII cost is $A_2=\sum_{t=1}^7f_2(t)$.

Now, we construct an embedded DTMC whose states are the EVs $E_i$, $i \in \mathcal{N}$, and subsequently an SMDP with the 5-tuple ($\mathcal{S},\mathbb{Z}_+,\rho,r,d$) described in Section~\ref{sec:smdp} as follows.
\begin{itemize}
    \item Embedded points are the states of the problem with the state space $\mathcal{S} = \{E_1,\ldots,E_N\}$.
    \item For each state $E_j$, the action is the value of the threshold $\tau_j \in \mathbb{Z}_+$, where $\mathbb{Z}_+$ denotes the set of positive integers, which constitutes the action space of the problem.
    \item There are two costs for this problem, which are the \emph{age penalty} cost, and the \emph{transmission} cost. For state $E_j$ and action $\tau_j$, we denote them with $a(E_j,\tau_j)=\mathbb{E}[A_j]$, and $c(E_j,\tau_j)=\mathbb{E}[C_j]$, respectively, and the expected total cost of the problem is denoted by $r(E_j,\tau_j)=a(E_j,\tau_j)+\lambda c(E_j,\tau_j)$. 
    \item Similarly, the expected duration for state $E_j$ for the same action equals $d(E_j,\tau_j)=\mathbb{E}[D_j]$.
    \item Lastly, $\rho(E_j,\tau_j,E_i)$ denotes the transition probability from state $E_j$ to the next state $E_i$ when action $\tau_j$ is applied.  
\end{itemize}
Once the parameters of the SMDP are found, it can be solved by using Algorithm~\ref{alg:SMDP}. In the next section, we will obtain the SMDP parameters by employing the DR-AMC and DR-DPH frameworks.

\section{Obtaining the SMDP Model} \label{sec:calculation}
In this section, we utilize the DR-AMC and DR-PH frameworks to obtain the SMDP parameters for the purpose of solving the unconstrained optimization problem given in \eqref{Opt1}. Then, we propose a method that adapts this solution to the constrained optimization problem given in \eqref{Opt2}. Specifically, we first model an embedded DTMC whose states are $E_i$, $i\in\mathcal{N}$, and then calculate the quantities $d(E_j,\tau_j)$, $a(E_j,\tau_j)$, and $\rho(E_i,\tau_j,E_j)$, for a given action $\tau_j$. Each cycle-$j$ starts with an EV $E_j$ which indicates $X_t=\hat{X}_t$, and it stays in synchronization for a duration $H_j$. When synchronized, no transmission is initiated, and no penalty occurs, thus the total cost is zero. The synchronization is broken with the state change of $X_t$ to any state $i\neq j$, and it lasts until it reaches an EP. We define the process $Y_j(t)\sim\text{DR-AMC} (\bm{\beta}_{j1},\tau_j,\bm\Theta,\bm{A}_{j1},\bm{A}_{j2},\bm{B}_{j1},\bm{B}_{j2})$ to represent the out-of-sync interval of cycle-$j$. Thus, the out-of-sync interval duration, denoted by $T_j$, has a DR-DPH distribution, i.e., 
$T_j\sim \text{DR-PH}(\bm{\beta}_{j1},\tau_j,\bm\Theta,\bm{A}_{j1},\bm{A}_{j2})$. Here, the transient state of the first regime corresponds to $i\in\mathcal{N}\backslash j$, enumerated as, 
\begin{align}
    \{1,\ldots,j-1,j+1,\ldots,N\},
\end{align}
and EVs $E_i$, $i\in \mathcal{N}$ are the absorption states. The first regime starts with the state transition of $j\to i$ of $X_t$ with probability  $q_{ji}$, $j\neq i$. Thus, we define the IPV for this regime with a row vector $\bm{\beta}_{j1}$, whose elements can be written as,
\begin{align}
    \{\bm{\beta}_{j1}\}_i= \dfrac{q_{ji}}{\sum_{k\neq j} q_{jk}}, \qquad i\neq j.
\end{align}
During this regime, the source does not initiate any transmission, thus only the absorbing state $E_j$ can be reached with a state change on $X_t$ to the estimated value $j$. This regime lasts until an absorption to $E_j$ occurs, or $t$ reaches the threshold $\tau$. Table~\ref{tab:Tprob1} provides the transition probabilities from the transient state $i$ from which the matrices $\bm{A}_{j1}$ and $\bm{B}_{j1}$ can be constructed. 
On the other hand, the transient states of the second regime are represented by the pair $(i,m)$, where $i\in\mathcal{N}\backslash j$ and $m\in\{1,\ldots,M\}$ correspond to the state of the source process and channel phase, respectively. We enumerate these states as follows,
\begin{align}
 \{(1,1),(1,2),\ldots,(j-1,M),(j+1,1),\ldots, (N,M)\}. 
\end{align}
Absorbing states in the second regime are the same as in the first regime. When the threshold is reached, the source transmits at each following time slot, and the first transmission starts at channel phase $k$ with probability $\gamma_k$. Therefore, the boundary transition matrix is obtained as,
\begin{align}
    \bm{\Theta}=\bm{I}_{N-1} \otimes \bm{\gamma}.
\end{align}
Different from the first regime, the process can be absorbed to $E_i$, $i\neq j$ if i) the state $X_t$ does not change, and ii) the transmission succeeds without preemption. These transition probabilities again are provided in Table~\ref{tab:Tprob2} from which the matrices $\bm{A}_{j2}$ and $\bm{B}_{j2}$ can be constructed similarly. 

\begin{table}[t]
    \caption{Transition probabilities for the process $Y_j(t)$ in the first regime.}
    \centering
    \begin{tabular}{|c|c|c|} 
    \hline
    \multicolumn{3}{|c|}{Transition probabilities from state $i$}\\ \hline
    To  & Condition & Probability\\ 
    \hline
    $i'$& $i'\neq i,j$ & $q_{ii'}$\\ \hline
    $E_j$& - &  $q_{ij}$  \\ \hline
    $E_{i}$&  $i'\neq j$ & $0$\\ \hline
    \end{tabular}
    \label{tab:Tprob1}
\end{table}

\begin{table}[t]
    \caption{Transition probabilities for the process $Y_j(t)$ in the second regime.}
    \centering
    \begin{tabular}{|c|c|c|} 
    \hline
    \multicolumn{3}{|c|}{Transition probabilities from state $(i,m)$}\\
    \hline
    To  & Condition & Probability\\ 
    \hline
    $(i,\ell)$& - & $q_{ii}g_{m\ell}$\\
    \hline
    $(i',\ell)$& $i'\neq i,j$ &$q_{ii'}\gamma_{\ell}$\\\hline
    $E_j$& - & $q_{ij}$  \\\hline
    $E_{i}$& $i'\neq j$& $(1-q_{ii})h_m$\\
    \hline
    \end{tabular}
    \label{tab:Tprob2}
\end{table}

\begin{example} \label{ex:cycle2}
    Consider a source process with three states, $N=3$, and the channel 
    modeled by DPH($\bm{\gamma},\bm{G}$) with two phases, $M=2$. For cycle-$2$ and given $\tau_2$, we construct the corresponding AMC $Y_2(t)\sim\text{DR-AMC}(\bm{\beta}_{21},\tau_2,\bm\Theta,\bm{A}_{21},\bm{A}_{22},\bm{B}_{21},\bm{B}_{22})$ as follows. 
    Transient states of the first regime are $\{1,3\}$. The process is initiated when synchronization is broken, resulting in a state transition of the source from state $2$ to one of these states where the IPV is given by,
    \begin{align}
        \bm{\beta}_{21}=\begin{bmatrix}
         \dfrac{q_{21}}{q_{21}+q_{23}} & \dfrac{q_{23}}{q_{21}+q_{23}}
    \end{bmatrix}.
    \end{align} The transition probabilities among the transient states are the same as the original process, thus, $\bm{A}_{11}$ is obtained by removing the $2$nd row and column of the $\bm{Q}$, 
    \begin{align}
        \bm{A}_{21}&=\begin{bmatrix}
        q_{11} & q_{13} \\ q_{31} & q_{33}
    \end{bmatrix}.
    \end{align} There are $N=3$ absorbing states $\{E_1,E_2,E_3\}$. However, since there is no transmission during this phase, only $E_2$ can be reached with probability $q_{i2}$ from transient state $i=1,3$, and $\bm{B}_{11}$ can thus be expressed as,
    \begin{align}
        \bm{B}_{21}=\begin{bmatrix} 0 & q_{12} & 0 \\ 0& q_{32} & 0
    \end{bmatrix}.
    \end{align} The transient states of the second regime are the pairs $(i,m)$ in the order of $\{(1,1),(1,2),(3,1),(3,2)\}$. At the start of the second regime, the probability of the source being in state $i$ can be obtained by $\bm{\beta}_1\bm{A}^{\tau-1}$. In addition, the channel process is initiated from phase $m$ with probability $\gamma_m$. Since the initialization of the channel process is independent from the source state, the IPV for this regime can be written as $\bm{\beta}_1\bm{A}^{\tau-1}\bm{\Theta}$ with
    \begin{align}
        \bm{\Theta}= \begin{bmatrix}
        \gamma_1 & \gamma_2 & 0 & 0\\
        0 & 0 & \gamma_1 & \gamma_2
        \end{bmatrix}.
    \end{align} 
    The transition probabilities among the transient states can be obtained from Table~\ref{tab:Tprob2} as
    \begin{align}
        \bm{A}_{22}&=\begin{bmatrix}
        q_{11} g_{11} & q_{11}g_{12} & \gamma_1 q_{13} & \gamma_2 q_{13} \\
        q_{11} g_{21} & q_{11}g_{22} & \gamma_1 q_{13} & \gamma_2 q_{13}\\
        \gamma_1 q_{31} & \gamma_2 q_{31}& q_{33} g_{11} &  q_{33} g_{12} \\
        \gamma_1 q_{31} & \gamma_2 q_{31} & q_{33} g_{11} &  q_{33} g_{12} 
        \end{bmatrix} \\
        &=\begin{bmatrix}
        q_{11}\bm{G} & \begin{matrix}
            q_{13} \bm{\gamma} \\  q_{13} \bm{\gamma}
        \end{matrix}   \\
        \begin{matrix}
            q_{31} \bm{\gamma} \\  q_{31} \bm{\gamma}
        \end{matrix} & q_{33}\bm{G}
        \end{bmatrix}
    \end{align} 
    Similar to the first regime, in this regime, the absorbing state $E_2$ can be reached with probability $q_{i2}$ from transient state $i=1,3$. On the other hand, the remaining absorbing states $E_i$, $i=1,3$ are reached from state $(i,m)$ with probability $q_{ii}h_m$, which is the product of the probabilities of the source staying in the same state and the successful transmission from channel phase $m$. Thus, APTS $\bm{B}_{22}$ can be written as,
    \begin{align}
        \bm{B}_{22}&=\begin{bmatrix}
        q_{11}h_1 & q_{12} & 0  \\
        q_{11}h_2 & q_{12} & 0 \\
        0 & q_{32} & q_{33}h_1 \\
        0 & q_{32} & q_{33}h_2
    \end{bmatrix}.
    \end{align}
\end{example}

\subsection{Derivation of $a(E_j,\tau_j)$}
For any given AoII penalty function $f_j(t)$, the expected age cost can be calculated from the distribution in \eqref{eq:pt2} as
\begin{align}
    a(E_j,\tau_j) &= \mathbb{E}\left[\sum_{t=1}^{T_j} f_j(t) \right]=\sum_{t=1}^{T_j} f_j(t)p_{T_j}(t). \label{eq:aj}
\end{align}
Next, we provide the closed-form expression for $a(E_j,\tau_j)$ when the AoII penalty functions $f_j(t)$ are polynomial functions of $t$. 

\begin{lemma}[Polynomial AoII Penalty Functions]
    If the AoII penalty function for estimation value $j$ is polynomial with degree $K_j$, i.e., 
    $f_{j}(t)=\sum_{k=0}^{K_j} w_{k,j} t^k$ where $w_{k,j}$ are the polynomial coefficients, then the 
    following closed-form expression holds for $a(E_j,\tau_j)$,
    \begin{align}
         \sum_{k=0}^{K_j} w_{k,j}\sum_{m=1}^k \dfrac{S(m+1,n+1)}{n+1} \mu_{T_j}(m). \label{closedform}
    \end{align}
\end{lemma}

\begin{Proof}
    The expression we want to calculate is
    \begin{align}
         \mathbb{E}\left[\sum_{t=1}^{T_j} f_j(t) \right]&=\mathbb{E}\left[\sum_{k=0}^{K_j} w_{k,j} \sum_{t=1}^{T_j} t^k\right ] \\
         &=\sum_{k=0}^{K_j} w_{k,j}\mathbb{E}\left[ \sum_{t=1}^{T_j} t^k\right ].
         \label{eq:ps1}
    \end{align}
    Using Faulhaber's formula \cite{bagui2024stirling}, the expected value of a finite power sum from $1$ to $T$ can be written in terms of the ordinary moments of the random variable $T$ 
    as follows,
    \begin{align}
        \mathbb{E}\left[\sum_{t=1}^{T}t^k\right]=\sum_{m=1}^k \dfrac{S(m+1,n+1)}{n+1} \mathbb{E}[T^m]. \label{eq:faul}
    \end{align}
    The proof is completed by using the relation between $\mu_{jm}$ and $\nu_{jm}$ in \eqref{eq:mom}, and inserting \eqref{eq:faul} into \eqref{eq:ps1}.
\end{Proof}

\subsection{Derivation of $d(E_j,\tau_j)$}
The expected duration of cycle-$j$, denoted by $d(E_j,\tau_j)$, is the sum of $\mathbb{E}[H_j]$ and $\mathbb{E}[T_j]$. The former term equals the expected number of trials until success with failure probability $q_{jj}$, which is $\mathbb{E}[H_j]=\frac{1}{q_{jj}}$. The second term can be calculated from \eqref{eq:pow} for $m=1$, which is written explicitly as
\begin{align}
  &  \sum_{t=1}^{\tau-1} t\bm{\beta}_{j1}\bm{A}_{j1}^{t-1}(\bm{1}-\bm{A}_{j1}\bm{1}) \nonumber\\ 
  &+\bm{\beta}_{j2}\left(\tau_j(\bm{I}-\bm{A}_{j2})^{-1}+\bm{A}_{j2}(\bm{I}-\bm{A}_{j2})^{-2} \right)(\bm{1}-\bm{A}_{j2}\bm{1}).  
\end{align}
\subsection{Derivation of $c(E_{j},\tau_j)$}
Since transmission continues each time slot in the second regime, the expected duration of the transmissions equals the number of transient states visited in the second regime. From the fundamental matrix definition in \eqref{eq:F}, it can be shown that
\begin{align}
    c(E_j,\tau_j)=\bm{\beta}_2(\bm{I}-\bm{A}_{j2})^{-1}\bm{1}.
\end{align}
\subsection{Derivation of $\rho(E_j,\tau_j,E_i)$}
Consider the absorbing states $j\neq i$ which can only occur in the second regime. For the threshold value $\tau_j$, its absorption probability can be calculated from \eqref{eq:prob_dr2} as
\begin{align}
   \rho(E_j,\tau_j,E_i)&=\bm{\beta}_{j2} (\bm{I}-\bm{A}_{j2})^{-1}\bm{B}_{j2}\bm{e}_i, \quad i\neq j.
\end{align}
Then, the self-transition probability for $E_j$ can be written as,
\begin{align}
    \rho(E_j,\tau_j,E_j)=1-\sum_{k\neq j}\rho(E_j,\tau_j,E_k).
\end{align}

\subsection{Solving the Constrained Problem}
SMDP formulation given above solves the unconstrained problem in \eqref{Opt1} for a given coefficient $\lambda$. In this part, we adopt this solution for the constrained problem of \eqref{Opt2}. It is known from \cite{makowski1986} that there exists a Lagrangian coefficient $\lambda^*$ such that the optimum policy obtained for the unconstrained problem is also optimum for the constrained problem either when (i) the constrained problem attains $TC^{\lambda^*}=\alpha$, or (ii) $\lambda=0$ and $TC^{0}\leq \alpha$, where $TC^\lambda$ is the time average of the transmission cost obtained from the unconstrained problem with coefficient $\lambda$. However, from the nature of the discrete-time system, a deterministic policy on the boundary of the constraint set may not exist. For such cases, a mixture of multiple deterministic policies can be used to obtain an optimal policy for the constrained problem. Among many methods, we adopt the non-randomized past-dependent policy, namely, \emph{steering algorithm} in \cite{ross1989randomized}. In the steering algorithm, we consider two policies corresponding to coefficients $\lambda_-$ and $\lambda_+$ such that $TC^{\lambda_-}\leq\alpha\leq TC^{\lambda_+}$, and the algorithm switches between the two policies based on the current sampling rate after each cycle. The general procedure of the algorithm is summarized in Algorithm~\ref{alg:steer} where $D^{(m)}$ and $C^{(m)}$ denote the duration and total number of transmission slots, respectively, in the $m$th cycle.

\begin{algorithm}[h]
    \caption{Steering algorithm}\label{alg:steer}
    \begin{algorithmic}
    \State \textbf{Input:} $\lambda_-$ and $\lambda_+$ such that $TC^{\lambda_-}\leq\alpha\leq TC^{\lambda_+}$
    \State \textbf{Initialize:} $TC_0=0$;
        \For{$m \gets 1$ to $M$}                    
    \If{$TC_{m-1}< \alpha$}
    \State Apply optimum thresholds obtained for $\lambda_+$; 
    \Else
    \State Apply optimum thresholds obtained for $\lambda_-$;
    \EndIf
    \State{$TC_m \gets \frac{\sum_{n=1}^m C^{(n)}}{\sum_{n=1}^m D^{(n)}}$};
    \EndFor     
    \end{algorithmic}
\end{algorithm}

\section{Numerical Results} \label{sec:numerical}
\begin{figure}
    \centering
    \includegraphics[width=0.60\linewidth]{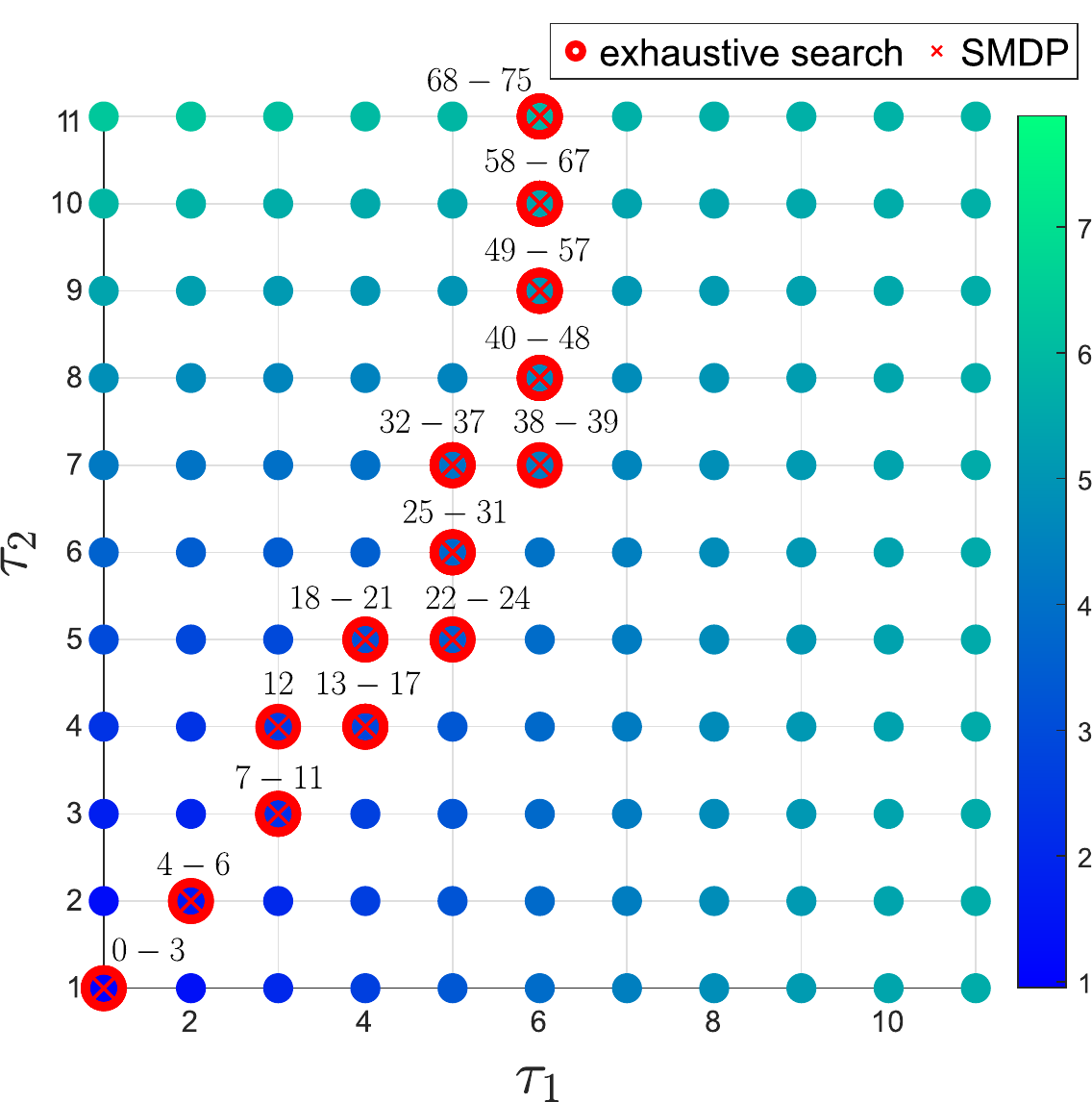}
    \caption{Each point shows the average AoII cost when using the threshold pair $(\tau_1,\tau_2)$ with a color map. Circle and cross markers are used to show the threshold values that minimize the average cost 
    using exhaustive search and the proposed SMDP algorithm, respectively, for a given weight $\lambda$, the values of which are given along with the markers.
    As an example, when $\lambda \in \{ 68,\ldots,75 \}$, the threshold pair $(6,11)$ is shown to be the optimum multi-threshold policy using both methods.}    \label{fig:binary}
\end{figure} 

In this section, we present numerical examples for validating the SMDP model of the paper along with comparisons with two benchmark policies: i) \emph{single threshold (ST) policy} for which the source waits for a single system-wide threshold $\tau$ while there is a mismatch between $X_t$ and $\hat{X}_t$, which is a widely used approach in the literature \cite{maatouk2020, chen2022preempting, maatouk2022age, bountrogiannis2024age}, and the value of $\tau$ which minimizes the overall cost is found by line search, ii) \emph{random sampling (RS) policy} \cite{cosandal_etal_TRIT24} in which transmission happens with probability $\xi$ at any time during the out-of-sync interval, and again the value of $\xi$ which minimizes the overall cost for RS is found by line search.

In the first example, $X_t$ has probability transition matrix $\bm{Q}_1$
\begin{align}
    \bm{Q}_1=\begin{bmatrix}
        0.65 & 0.35 \\ 0.25 & 0.75
    \end{bmatrix}, 
\end{align}
under the AoII penalty functions 
\begin{align}
    f_1(x)=x^2+\frac{1}{2}x+\frac{1}{3}, \quad f_2(x)=\frac{7}{10}x^2+\frac{3}{5}x+\frac{1}{2},
\end{align}
and we consider a geometrically distributed channel delay with parameter $0.8$. Fig.~\ref{fig:binary} illustrates the overall cost for each threshold pair $(\tau_1,\tau_2)$ (obtained analytically) illustrated with a color map in Fig.~\ref{fig:binary}. For each integer $\lambda$ between $0$ and $75$, the threshold pair that minimizes \eqref{Opt1} is obtained with exhaustive search, and shown with a red circle. In addition, the optimum threshold pair obtained by the proposed SMDP method is marked with a cross. The perfect match between these marks verifies the optimality of our algorithm.

In the second numerical example, we study two DTMC information sources. 
The first 3-state source has a probability transition matrix $\bm{Q}_2$
\begin{align}
    \bm{Q}_2= \begin{bmatrix}
        0.7 & 0.2 & 0.1 \\ 0.3 & 0.6 & 0.1 \\ 0.2 & 0.3 & 0.5
    \end{bmatrix},
\end{align}
and we employ AoII penalty functions 
\begin{align}
    f_1(x)=x^2+\frac{1}{2}, \ f_2(x)=\frac{1}{2}x^2+\frac{1}{2}x, \ f_3(x)=\frac{1}{3}x^2+\frac{1}{4}. 
\end{align}
The second source with $N=10$ states has a probability transition matrix $\bm{Q}_3$ whose diagonal elements are linearly spread in the interval $[0.4,0.6]$, and similarly, off-diagonal elements are linearly spread in the interval $[0.5\frac{1-q_{nn}}{N-1}, 1.5\frac{1-q_{nn}}{N-1}]$. For this source, we consider the AoII penalty functions $f_n(x)=\frac{1}{n}x^2+\frac{1}{N+1-n}x$ for state $n$. We use the same channel model as before. Then, for a given weight parameter $\lambda$, the proposed SMDP algorithm obtains the multi-threshold optimum policy. The average cost obtained with the SMDP algorithm along with the RS and ST policies is depicted in Fig.~\ref{fig:gen}. For ST and SMDP policies, both analytical and simulation results are obtained, and the strong agreement between them verifies our analytical results. Since all AoII penalty functions are polynomials in the examples, we employed the closed-form expression \eqref{closedform} for finding the parameters of the SMDP model. On the other hand, only simulation results are used for RS. We observe that our proposed SMDP-based policy outperforms both benchmark policies substantially, and all converge to the same policy when $\lambda \rightarrow 0$ which corresponds to the {\em always transmit} policy. We additionally observe that the performance gain increases with $\lambda$, for instance, average cost of SMDP at $\lambda=4$ for the process $\bm{Q}_3$ is around 25\% less than ST policy, and 30\% less than PS policy.
 
\begin{figure}
    \begin{center}
    \subfigure[]{\includegraphics[width=0.85\linewidth]{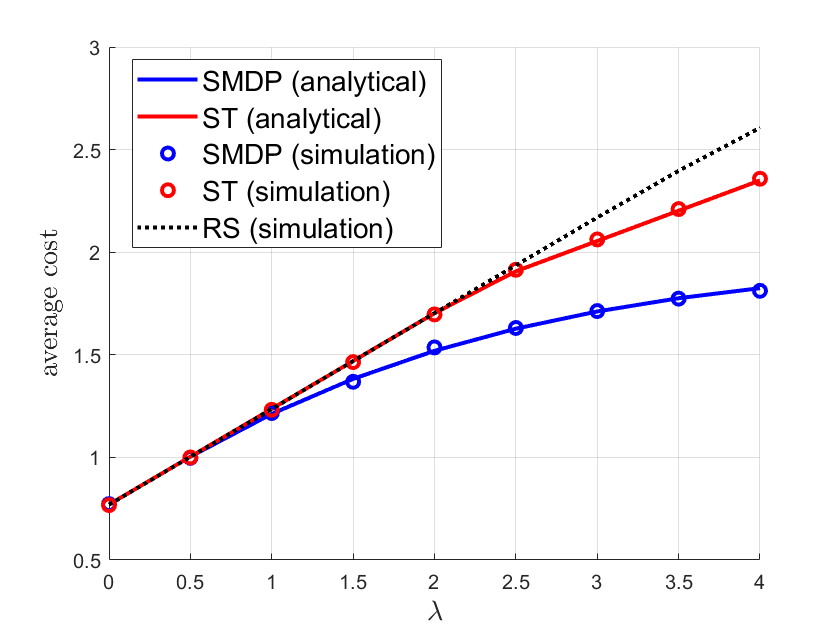}} ~ 
    \subfigure[]{\includegraphics[width=0.85\linewidth]{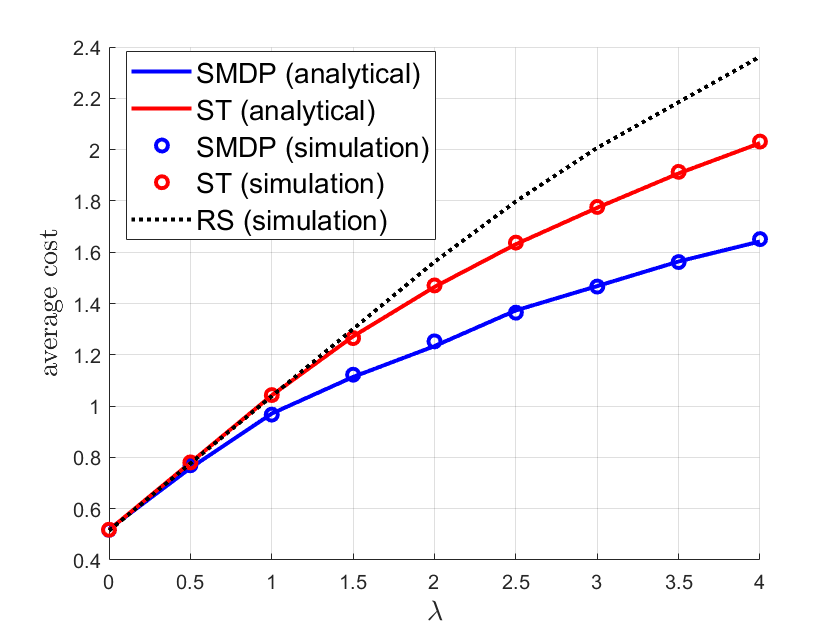}} 
    \end{center}  
    \caption{Comparison of benchmark policies with proposed SMDP policy with varying $\lambda$ for a) $\bm{Q}_2$ $(N=3)$, and b) $\bm{Q}_3$ $(N=10)$.}
    \label{fig:gen}
\end{figure}

\begin{figure}
    \begin{center}
    \includegraphics[width=0.85\linewidth]{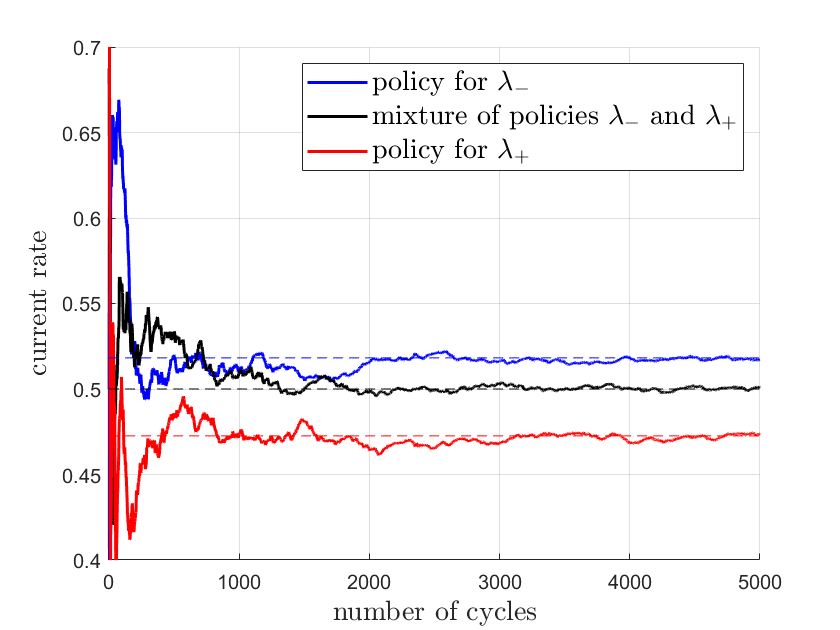}
    \end{center}  
    \caption{Applying the steering algorithm for $\lambda_{-}=1.134$, $\lambda_{+}=1.135$, $\bm{Q}_4$ and $\alpha=0.5$, and comparing with the deterministic policies. The long-term sampling rates are denoted with dashed lines.} 
    \label{fig:steer}
\end{figure}

In the final example, we study a two-phase channel delay distributed according to DPH($\gamma,\bm{G}$) with parameters,
\begin{align}
    \bm{\gamma}=\begin{bmatrix}
        1 & 0 \\ 
    \end{bmatrix}, \quad \bm{G}=\begin{bmatrix}
        0.7 & 0.2 \\ 0.1 & 0.6
    \end{bmatrix}, \label{eq:2gc}
\end{align}
for the process 
\begin{align}
    \bm{Q}_4=\begin{bmatrix}
        0.6 & 0.25 & 0.15 \\
        0.25 & 0.55 & 0.2 \\
        0.2 & 0.3 & 0.5
        \end{bmatrix},
\end{align}
and the linear AoII penalty functions
\begin{align}
    f_1(x)=x+\dfrac{1}{2}, \quad f_2(x)=\frac{1}{2}x+1, \quad f_3(x)=\frac{1}{3}x+\frac{1}{4}.
\end{align}
The results are depicted in Fig.~\ref{fig:gen} for which we observe trends between benchmark policies and the optimal policy, similar to what we have obtained before. Finally, for the same setting, we solve the constrained problem in \eqref{Opt2} for $\alpha=0.5$. We obtain coefficients $\lambda_{-}=1.134$ and $\lambda_{+}=1.135$ with the corresponding sampling rates $TC^{-}=0.5182$, $TC^+=0.4726$, and optimum threshold values $\bm{\tau}_{-}=\begin{bmatrix}
    1 & 1 & 4
\end{bmatrix}$, $\bm{\tau}_{+}=\begin{bmatrix}
    1 & 1 & 5
\end{bmatrix}$. Fig.~\ref{fig:steer} illustrates the simulation results for the mixture of these policies obtained by the steering algorithm in Alg.~\ref{alg:steer}. We observe that the transmission cost of the mixture policy converges to the budget at a similar rate to the deterministic policies obtained for the coefficients $\lambda_{-}$ and $\lambda_{+}$.

\begin{figure}
    \begin{center}
    \includegraphics[width=0.85\linewidth]{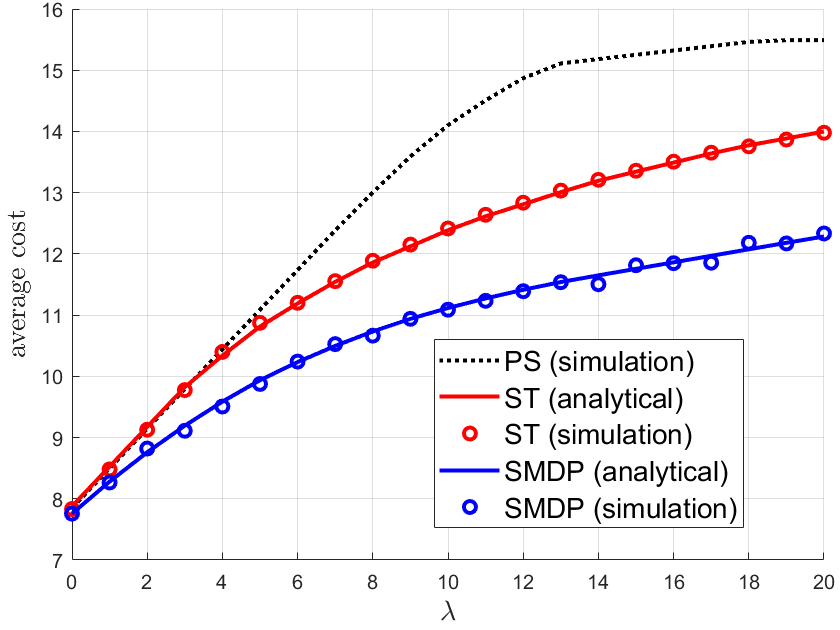}
    \end{center}  
    \caption{Comparison of benchmark policies with proposed SMDP policy with varying $\lambda$ for $\bm{Q}_4$ under the two-phase channel DPH($\bm{\gamma},\bm{G}$), which is defined in \eqref{eq:2gc}.}
    \label{fig:gen}
\end{figure}

\section{Conclusions} \label{sec:conclusions}
We solve the cost minimization problem for a push-based remote estimation system that consists of a DTMC source process, a monitor, and a forward delay channel modeled by a DPH distribution. In the proposed setting, the source initiates a transmission when the AoII value exceeds an estimation-based threshold that is to be obtained for each estimation to solve the cost minimization problem. Initiated transmission lasts until it is successful, or it is preempted, and a new transmission is initiated if the source process changes. 

We formulate the problem as an SMDP with the same state-space as the original DTMC process, using the embedded DTMC approach. To obtain the parameterization of the SMDP, we propose to utilize the DR-AMC and DR-DPH frameworks that can further be adapted to analyze other freshness metrics for threshold-based policies. Cost of the problem is defined as the weighted combination of transmission cost and any estimation-based functions of AoII. In addition, closed-form expressions are obtained for polynomial functions. Furthermore, a mixture of policy methods, namely the steering algorithm, is adapted to solve the constrained problem. Analytical methods and optimality are verified by numerical results by comparing the proposed approach against exhaustive search and benchmark policies. Results from the exhaustive search verify the optimality of our proposed algorithm. We additionally observe that the performance gain in using the proposed multi-threshold approach against the single threshold and random sampling policies increases with increasing cost of the transmission, which makes it advantageous for settings where the sampling is rare because of low energy availability.

\appendices

\section{Examples of DPH distributions} \label{app:channel}
Here, we obtain the DPH representation of the geometric distribution, mixture of geometric distributions, and bounded support distributions, which are used to model channel delays in this paper.

\paragraph{Geometric distribution} 
When the channel delay $D$ is distributed according to ${Geo}(\theta)$, 
at each slot, the transmission is complete with probability $\theta$, and fails to complete with probability $1-\theta$. In other words, we have the representation $D \sim \text{DPH}(\bm{\gamma},\bm{G})$ where we have a single phase, and
\begin{align}
    \bm{\gamma}=1, \quad   \bm{G}=1-\theta.
\end{align}

\paragraph{Mixture of geometric distribution (MGD)}
There are $K$ geometric distributions ${Geo}(\theta_k)$ with mixture weights 
$w_k$, for $k=1,\dots,K$. Channel delay $D$ is said to have an MGD distribution 
characterized by the parameters $\{\theta_k\}_{k=1}^K$ and $\{w_k\}_{i=1}^K$ when the delay $D$ is geometrically distributed with parameter $\theta_k$ with probability $w_k$.  
For each transmission, the channel is in one of the $K$ conditions, and the weights represent the probability of experiencing the condition. The representation $D \sim \text{DPH}(\bm{\gamma},\bm{G})$ has $K$ phases where
\begin{align}
    \bm{\gamma}&=\begin{bmatrix}
        w_1 & \cdots & w_K
    \end{bmatrix}, \\
    \bm{G}&=\begin{bmatrix}
        1-\theta_1 &  0 & \cdots  & 0 \\
        0 & 1-\theta_2   & \cdots & 0 \\
        \vdots & &\ddots & & \\
        0 & \cdots & & 1-\theta_K  
    \end{bmatrix}.
\end{align}

\paragraph{Bounded support distribution}
Consider a channel delay $D$ described by a finite probability vector $\bm{\zeta}$, i.e., 
transmission duration is $k$ slots with probability $\zeta_k$ for $k \leq P$. 
The representation $D \sim \text{DPH}(\bm{\gamma},\bm{G})$ has $P$ phases where
\begin{align}
 \bm{\gamma}&=\begin{bmatrix}
    1 & \cdots & 0
\end{bmatrix}, \\
    \bm{G} 
    &=\begin{bmatrix}
        0 & 1-\frac{\zeta_1}{\sum_{p=1}^P\zeta_p} & 0 & \cdots & 0 \\
        0 & 0 & 1-\frac{\zeta_2}{\sum_{p=2}^P\zeta_p} &  \cdots & 0 \\
        \vdots & & & & \vdots \\
        0 & \cdots & & & 1-\frac{\zeta_{P-1}}{\sum_{p=P-1}^P\zeta_p} \\
        0 & \cdots & & & 0
    \end{bmatrix}.
\end{align}

\section{Moments of DR-DPH Distributions} \label{app:fact}
The $m$th factorial moment of $T$ distributed according to $\text{DR-DPH}(\tau,\bm{\beta},\bm{\Theta},\bm{A}_1,\bm{A}_2,\bm{B}_1,\bm{B}_2)$ is equivalent to
\begin{align}
    \nu_T(i)=&\sum_{t=1}^{\tau-1}t^{\underline{m}}\bm{\beta}_{1}\bm{A}_{1}^{t-1}(\bm{1}-\bm{A}_{1}\bm{1})\nonumber\\ &+\underbrace{\sum_{t=\tau}^{\infty}t^{\underline{m}}\bm{\beta}_{2}\bm{A}_{2}^{t-\tau}(\bm{1}-\bm{A}_{2}\bm{1})}_{I_m}. \label{proof_nu}
\end{align}
The first term is a finite sum which can be calculated for finite $\tau$. The term $I_m$, on the other hand, includes an infinite summation, and special transformations are required to obtain an expression for it in closed-form. We rewrite this term as
\begin{align}
    I_m  &=\bm{\beta}_{2} \left(\sum_{t=\tau}^{\infty}t^{\underline{m}}\bm{A}_{2}^{t-\tau}\right)(\bm{1}-\bm{A}_{2}\bm{1}). \label{eq:Im}
\end{align}
In order to evaluate the sum term $\sum_{t=\tau}^{\infty}t^{\underline{m}}\bm{A}_{2}^{t-\tau}$, we utilize the following identity for falling factorials \cite{graham1994concrete},
\begin{align}
    x^m&=\sum_{n=1}^m S(m,n)x^{\underline{m}}, \label{fall1} \\
    (a+b)^{\underline{m}}&=\sum_{n=0}^m \binom{m}{n}a^{\underline{m-n}}b^{\underline{n}}, \label{fall2} 
\end{align}
where $S(m,n)$ is the second kind Stirling number  whose values for $m,n\leq4$ are presented in Table~\ref{tab:Stirling}. In addition, from \eqref{factorialMoments}, one can obtain the identity
\begin{align}
    \sum_{n=0}^\infty n^{\underline{m}}\bm{A}^{n}=&m!\bm{A}^m (\bm{I}-\bm{A})^{-m-1}. \label{fall3}
\end{align}
We now have two separate cases as $\tau\geq m$ and $\tau<m$.

\paragraph{Case $\tau\geq m $}
By changing variable $d=t-\tau$ and applying identity \eqref{fall2}, we get
\begin{align}
    \sum_{t=\tau}^{\infty}t^{\underline{m}}\bm{A}_{2}^{t-\tau}&=\sum_{d=0}^{\infty}(t+\tau)^{\underline{m}}\bm{A}_{2}^{d} \label{eq:c1s0} \\
    &= \sum_{d=0}^{\infty} \sum_{r=0}^m \binom{m}{r} \tau^{m-r} d^r \bm{A}_{2}^{d }  \label{eq:c1s1} \\
    &=\sum_{r=0}^m \binom{m}{r} \tau^{m-r}  \sum_{d=0}^{\infty} d^r \bm{A}_{2}^{d} \label{eq:c1s2} \\
    &=\sum_{r=0}^m \binom{m}{r} \tau^{m-r} r!\bm{A}_2^r(\bm{I}-\bm{A}_2)^{-r-1}. \label{eq:c1}
\end{align}
We obtain \eqref{eq:c1s1} and \eqref{eq:c1} by applying \eqref{fall2} and \eqref{fall3}, respectively. Finally, we end up with $\nu_m$ for $\tau\geq m $ by inserting \eqref{eq:c1} into \eqref{eq:Im} and \eqref{proof_nu}, subsequently.

\paragraph{Case $\tau<m$}
For this case, we need to divide the sum term to avoid $\tau^{\underline{k}}$ for any $k>\tau$ as 
\begin{align}
    \sum_{t=\tau}^{\infty}t^{\underline{m}}\bm{A}_{2}^{t-\tau}=\sum_{t=\tau}^{m-1}t^{\underline{m}}\bm{A}_{2}^{t-\tau}+\sum_{t=m}^{\infty}t^{\underline{m}}\bm{A}_{2}^{t-\tau}.
\end{align}
Again, the finite sum can be calculated efficiently, and applying the previous steps to the infinite sum, we get
\begin{align}
    \sum_{t=m}^{\infty}t^{\underline{m}}\bm{A}_{2}^{t-\tau}=\sum_{r=0}^m \binom{m}{r} m^{\underline{m-r}} r!\bm{A}_2^r(\bm{I}-\bm{A}_2)^{-r-1} \bm{A}_2^{m-\tau}.
\end{align}
We conclude the proof by applying the identity \eqref{fall1} to obtain ordinary moments from the factorial moments as
\begin{align}
    \mu_T(m)=\sum_{r=0}^m S(m,r) \nu_T(r). \label{eq:mom_app}
\end{align}

\begin{table}[!t]
    \centering
    \caption{Stirling numbers of the second kind \( S(n, m) \) for \( n, m \leq 4 \).}
    \begin{tabular}{c|cccc}
        \hline
        $n \backslash m$ & 1 & 2 & 3 & 4 \\
        \hline
        1 & 1 &  &  &  \\
        2 & 1 & 1 &  &  \\
        3 & 1 & 3 & 1 &  \\
        4 & 1 & 7 & 6 & 1 \\
        \hline
    \end{tabular}
    \label{tab:Stirling}
\end{table}
    
\section{Standard MDP Formulation} \label{app:stand}
The conventional way to obtain the optimum policy is to define an MDP with the 5-tuple $(\mathcal{S},\mathcal{A},c,\rho)$, which is a special case of the SMDP formulation in Section~\ref{sec:smdp} with unit sojourn times.
First, we denote the \emph{channel phase process} by $P_t$, where $P_t\in \mathcal{M}=\{1,\dots M\}$ which is defined as follows. When the channel is not actively used (if no transmission is initiated yet, or the packet is just preempted with a state change on the source), $P_t=m$ with probability $\gamma_m$. During the transmission of a packet, it evolves from $P_t=m$ to $P_{t+1}=m'$ with probability $g_{mm'}$, or transmission succeeds with probability $h_m$.

The states of the problem are the values of the four-dimensional joint process $s_t=(X_t, \ P_t, \ \hat{X}_t, \ \text{AoII}_t)$, thus the state-space of this problem is $\mathcal{S}=\{\mathcal{N}\times \mathcal{M} \times \mathcal{N} \times \mathbb{N}\}$, where $\mathbb{N}$ is the set of non-negative integers. The action space of this MDP is $\mathcal{A}=\{0,1\}$, and the action is $\delta_t\in\mathcal{A}$. The cost function can be defined as $r(s_t,\delta_t)=\text{AoII}_t+\lambda\delta_t$. The formulation can be finalized with the transition probability function $\rho(s,a,s')$ for $s=(i,m,j,k)$, $\delta=u$, and $s=(i',m',j',k')$ as
\begin{align}
    \begin{cases}
        \gamma_{m'}q_{ii'}, &  u=0, \ i'\neq i, \  i\neq j, \ j'=j \ k'=k+1, \\
        \gamma_{m'}q_{ii'}, &   u=0, \ i'=j\neq i, \ j'=j, \ k'=0, \\
        \gamma_{m'}q_{ii'}, &  u=1, \ i'\neq i, \  i\neq j, \ j'=j  \ k'=k+1, \\
        g_{mm'}q_{ii}, & u=1, \ i'= i, \  i\neq j, \ j'=j,  \  k'=k+1, \\
        h_m\gamma_{m'}q_{ii}, & u=1, \ i'= i, \  i\neq j, \ j'=i  \  k'=0.
    \end{cases}
\end{align}
Notice that the state-space of this formulation includes infinitely many elements since AoII$_t$ can take any non-negative value. Therefore, in order to obtain the optimum policy using dynamic programming, the AoII$_t$ process should be truncated, giving rise to deviation from the optimality unless truncation is done properly.

\bibliographystyle{IEEEtran}
\bibliography{bibl}

\end{document}